\documentclass[12pt,a4paper,onecolumn]{article}

\usepackage[latin1]{inputenc}
\usepackage{amsmath}
\usepackage{amsfonts}
\usepackage{amssymb}
\usepackage{makeidx}
\usepackage{epsfig}
\usepackage{wrapfig}
\usepackage[english]{babel}
\usepackage[usenames,dvipsnames]{color}
\usepackage{colortbl, multirow}
\usepackage{subfig}
\usepackage{hyperref}

\usepackage[english]{varioref}

\definecolor{light-green}{rgb}{.75,1,.75}
\definecolor{light-red}{rgb}{1,.75,.75}

\author{Federico Garzarelli$^1$, Matthieu Cristelli$^{1,2,*}$, Andrea Zaccaria$^{1,2}$,\\ Luciano Pietronero$^{1,2}$\\\\
$^1$ \small Dip. di Fisica, Universit\`a ``Sapienza'', P.le A. Moro 2, 00185, Roma, Italy\\
$^2$ \small  ISC-CNR, Via dei Taurini 19, 00185, Roma, Italy\\
$^*$ \small Corresponding author: matthieu.cristelli@roma1.infn.it}
\title{Memory effects in stock price dynamics: evidences of technical trading}
\makeindex

\begin{document}
\maketitle

\begin{abstract}
Technical trading represents a class of investment strategies for Financial Markets
based on the analysis of trends and recurrent patterns of price time series. According standard 
economical theories these strategies should not be used because they cannot be profitable.
On the contrary it is well-known that technical traders exist and
operate on different time scales. In this paper we investigate if
technical trading produces detectable signals in price time series and if some kind of memory effect 
is introduced in the price dynamics. In particular we focus
on a specific \textit{figure} called \textit{supports and resistances}. We first develop a
criterion to detect the potential values of supports and resistances. As a second step,
we show that memory effects in the price dynamics are associated to these selected values. In fact we show
that prices more likely re-bounce than cross these values. Such an effect is a quantitative evidence 
of the so-called \textit{self-fulfilling prophecy} that is the self-reinforcement of agents' belief and sentiment
about future stock prices' behavior.
\end{abstract}

\section{Introduction}
Physics and mathematical methods 
derived from Complex Systems Theory and Statistical Physics 
have been shown to be effective tools  \cite{golub2010data,mitchell,vesp1,lazer2009life}  to provide a quantitative description 
and an explanation of many social \cite{goel2010consumer,twitter,ginzberg2009epidemics} 
and economical phenomena \cite{choi2009trends,miotrends,saavedra2011traders}.

In the last two decades Financial Markets have appeared as natural candidates for this interdisciplinary
application of methods deriving from Physics because a systematic approach to the issues set by this 
field can be undertaken. In fact
since twenty years there exists a very huge amount of high frequency data from stock
exchange which permit to perform experimental procedures as in Natural Sciences.
Therefore Financial Markets appear as a perfect playground where models and theories can be tested and
where repeatability of empirical evidences is a well-established feature differently
from, for instance, Macro-Economy and Micro-Economy. In addition the methods
of Physics have proved to be very effective in this field and have often given rise to
concrete (and profitable) financial applications.

The major contributions of Physics to the comprehension of Financial Markets on
one hand are focused on the analysis of financial time series' properties and on
the other hand on agent-based modeling \cite{abergel,varenna}. The former contribution provides
fundamental insights in the non trivial nature of the stochastic process performed
by stock price \cite{revBOU,Bouchaud2,farmerlongI} and in the role of the dynamic interplay between agents to
explain the behavior of the order impact on prices \cite{Bouchaud2,Bouchaud1,Bouchaud3,lillocoll,farmer6,weber1,ob_pil}.
The latter approach instead has tried to overcome the traditional economical models
based on concepts like price equilibrium and homogeneity of agents in order to
investigate the role of heterogeneity of agents and strategies with respect to the price dynamics
\cite{varenna,alfisoi,reviu,LMnature,giard_bou,caldase,paperoI,paperoII}.

In this paper we focus our attention on technical trading. 
A puzzling issue of standard economical theory of Financial Markets is that the strategies based 
on the analysis of trends and recurrent patterns 
(i.e. known indeed as technical trading or chartist strategies) should
not be used if all agents were rational because prices should
follow their fundamental values \cite{bucha,kreps} and no arbitrage opportunities should be present . Consequently these speculative
strategies cannot be profitable in such a scenario.\\
It is instead well-known that chartists (i.e. technical traders) exist and
operate on different time scales ranging from seconds to months.

In this paper we investigate if a specific chartist
strategy produces
a measurable effect on the statistical properties of price time series that is if there exist special values on which prices tend to bounce. As we are going
to see the first task that we must address consists in the formalization of this strategy in a suitable mathematical framework.\\
Once a 
quantitative criterion to select potential supports and resistances is developed, we investigate
if these selected values introduce memory effects in price evolution.\\
We observe that: i) the probability of re-bouncing on these selected values is higher than what expected and ii) the more the number of bounces on
these values increases, the more the probability of bouncing on them is high. In terms of 
agents' sentiment we can say that the more
 agents observe bounces the more they expect that the price will again bounce
on that value and their beliefs introduce a positive feedback which in turn reinforces the support or the resistance.
 
Moving to the paper organization, in the remaining of this section we give a brief introduction to technical trading.
\\In section 2 we introduce a quantitative criterion to select the price values which are 
potential supports or resistances.
\\In section 3 we show that the probability of bouncing on these selected values is higher 
than expected highlighting a detectable memory effect.\\
In section 4 we discuss the origin of such an effect and we assess the issue of which mechanisms can produces it.\\
In section 5 we give a statistical characterization of the main features of the pattern described 
by the price when it bounces on supports and resistances.\\
In section 6 we draw the conclusion of our work.

\subsection{The classical and the technical approaches} 
The classical approach in the study of the market dynamics is to build a stochastic model for the price dynamics with the so called martingale property $ E(x_{t+1} | x_{t}, x_{t-1}, \dots, x_{0}) = x_{t} \quad \forall t$ \cite{hull,Bouchaud-Potters,voit,Mantegna-Stanley}. The use of a martingale for the description of the price dynamics naturally arises from the hypothesis of efficient market and from the empirical evidence of the absence of simple autocorrelation between price increments. The consequence of this kind of model for the price is that is impossible to extract any information on the future price increments from an analysis of the past increments.

The technical analysis is the study of the market behavior underpinned on the inspection of the price graphs. The technical analysis permits the speculation of the future value of the price. According to the technical approach the analysis of the past prices can lead to the forecast of the future value of prices. This approach is based upon three basic assumptions \cite{tech}:
\begin{itemize}
\item[1] \textbf{the market discounts everything}: the price reflects all the possible causes of the price movements (investors' psychology, political contingencies and so on) so the price graph is the only tool to be considered in order to make a prevision.
\item[2] \textbf{price moves in trends}: price moves as a part of a trend, which can have three direction: up, down, sideways. According to the technical approach, a trend is more likely to continue than to stop. The ultimate goal of the technical analysis is to spot a trend in its early stage and to exploit it investing in its direction. 
\item[3] \textbf{history repeats itself}: Thousands of price graphs of the past have been analyzed and some figures (or patterns) of the price graphs have been linked to an upward or downward trend \cite{tech}. The technical analysis argues that a price trend reflects the market psychology. The hypothesis of the technical analysis is that if these patterns anticipated a specific trend in the past they would do the same in the future. The psychology of the investors do not change over time therefore an investor would always react in the same way when he undergoes the same conditions.
\end{itemize}

One reason for the technical analysis to work could be the existence of a feedback effect called \textit{self-fulfilling prophecy}. Financial markets have a unique feature: the study of the market affects the market itself because the results of the studies will be probably used in the decision processes by the investors\footnote{Other disciplines such as physics do not have to face this issue.}. The spread of the technical analysis entails that a large number of investors have become familiar with the use of the so called \textit{figures}. A figure is a specific pattern of the price associated to a future bullish or bearish trend. Therefore, it is believed that a large amount of money have been moved in reply to bullish or bearish \textit{figures} causing price changes. In a market, if a large number of investors has the same expectations on the future value of the price and they react in the same way to this expectation they will operate in such a way to fulfill their own expectations. As a consequence, the theories that predicted those expectation will gain investors' trust triggering a positive feedback loop. In this paper we tried to measure quantitatively the trust on one of the figures of technical analysis.

\subsection{Supports and Resistances}
Let us now describe a particular figure: supports and resistances. The definition of support and resistance of the technical analysis is rather qualitative: a support is a price level, local \textit{minimum} of the price, where the price will bounce on other times afterward while a resistance is a price level, local \textit{maximum} of the price, where the price will bounce on other times afterward. We expect that when a substantial number of investors detect a support or a resistance the probability that the price bounces on the support or resistance level is bigger than the probability the price crosses the support or resistance level. Whether the investors regard a local minimum or maximum as a support or a resistance or not  can be related to: i) the number of previous bounces on a given price level, ii) the time scale. The investors could \textit{a priori} look at heterogeneous time scales.
This introduces two parameters which we allow to vary during the analysis in order to understand if and how they affect our results.


\section{Supports and Resistances: quantitative definition}
\begin{figure}[h!]
\centering
\includegraphics[scale=0.5]{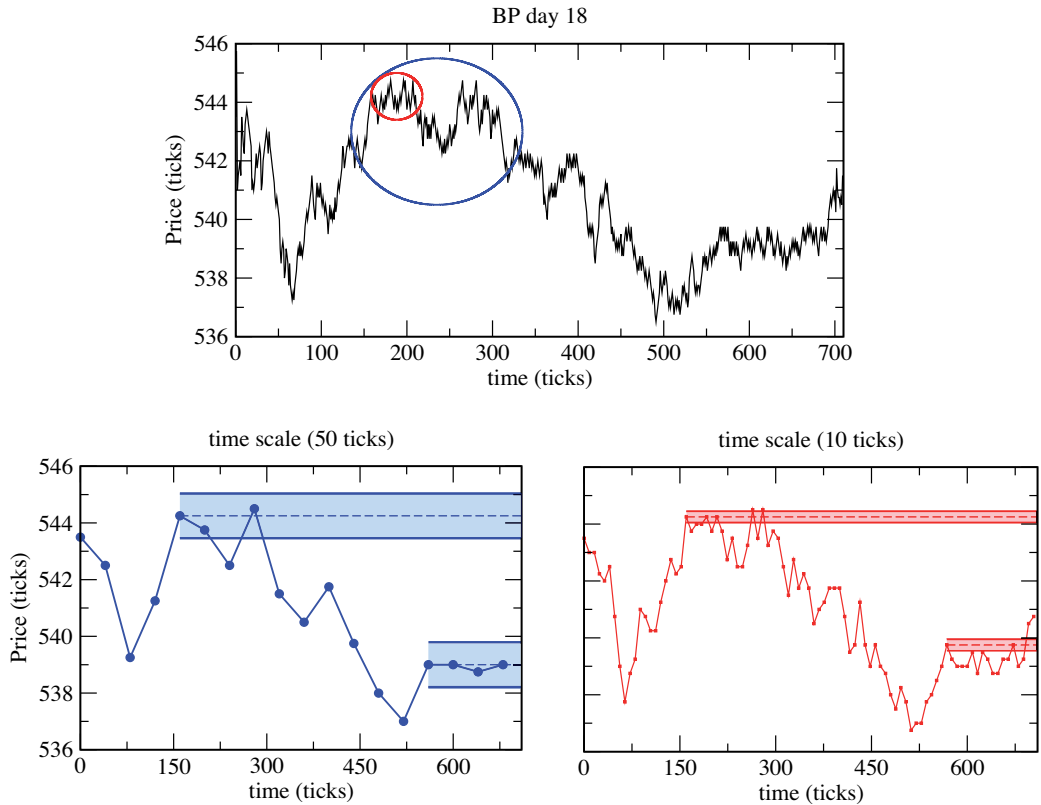} 
\caption{The graph above illustrates the price (in black) \textit{tick-by-tick} of the stock British Petroleum in the 18th trading day of 2002. The blue and red circles define two regions of different size where we want to look for supports and resistances. The graph below in the left shows the price (in blue) extracted from the time series \textit{tick-by-tick} picking out a price every 50 ticks in the same trading day of the same stock. The graph below in the right shows the price (in red) extracted from the time series \textit{tick-by-tick} picking out a price every 10 ticks. The horizontal lines represent the stripe of the resistance to be analyzed.}
\label{fig:bpsuppres}
\end{figure}

 One has to face two issues trying to build a quantitative definition of support and resistance:

\paragraph*{1} We define a bounce of the price on a support/resistance level as the event of a future price entering in a stripe centered on the support/resistance and exiting from the stripe without crossing it.
Furthermore, we want to develop a quantitative definition compatible with the way the investors use to spot the support/resistances in the price graphs. In fact the assumed memory effect of the price stems from the visual inspection of the graphs that comes before an investment decision. To clarify this point let us consider the three price graphs in fig. \ref{fig:bpsuppres}. The graph in the top panel shows the price tick-by-tick of British Petroleum in the 18th trading day of 2002. If we look to the price at the time scale of the blue circle we can state that there are two bounces on a resistance, neglecting the price fluctuation in minor time scales. Conversely, if we look to the price at the time scale of the red circle we can state that there are three bounces on a resistance, neglecting the price fluctuation in greater time scales. The bare eye distinguishes between bounces at different time scales. Therefore we choose to analyze separately the price bounces at different time scales.
To select the time scale to be used for the analysis of the bounces, we considered the time series $P_{\tau}(t_i)$ obtained  picking out a price every $\tau$ ticks from the time series \textit{tick-by-tick}\footnote{We have a record of the price for every operation.}. The obtained time series is a subset of the original one: if the latter has $N$ terms then the former has $[N/\tau]$ terms\footnote{The square brackets $[\,\,]$ indicate the floor function defined as  $[ x ] =\max\, \{m\in\mathbb{Z}\mid m\le x\}$}. In this way we can remove the information on the price fluctuations for time scales less than $\tau$. The two graphs in fig.\ref{fig:bpsuppres} (bottom panel) show the price time series obtained from the \textit{tick-by-tick} recordings respectively every 50 and 10 ticks. We can see that the blue graph on the left shows only the bounces at the greater time scale (the time scale of the blue circle) as the price fluctuations at the minor time scale (the one of the red circle) are absent. Conversely these price fluctuations at the minor time scale are evident in the red graph on the right. 

\paragraph*{2}The width $\delta$ of the stripe centered on the support or resistance at the time scale $\tau$ is defined as
 \begin{equation}
\delta(\tau) = \left[ \frac{\tau}{N}\right] \sum_{k=1}^{ \left[ \frac{\tau}{N}\right]-1}|P_{\tau}(t_{k+1})-P_{\tau}(t_k)|
\end{equation}
that is the average of the absolute value of the price increments at time scale $\tau$.
Therefore $\delta$ depends both on the trading day and on the time scale and it generally rises as $\tau$ does. In fact it approximately holds that $\delta_\tau \sim \tau^\alpha$ where $\alpha$ is the diffusion exponent of the price in the day considered. The width of the stripe represents the tolerance of the investors on a given support or resistance: if the price drops below this threshold the investors regard the support or resistance as broken. 

To sum up, we try to separate the analysis of the bounces of price on supports and resistances for different time scales. Provided this quantitative definition of support and resistance in term of bounces we perform an analysis of the bounces in order to determine if there is a memory effect on the price dynamics on the previous bounces and if this effect is statistically meaningful.

\section{Empirical evidence of memory effects}
The analysis presented in this paper are carried out on the high frequency (\textit{tick-by-tick}) time series of the price of 9 stocks of the London Stock Exchange in 2002, that is 251 trading days.
The analyzed stocks are: AstraZeNeca (AZN), British Petroleum (BP), GlaxoSmithKline (GSK), Heritage Financial Group (HBOS), Royal Bank of Scotland Group (RBS), Rio Tinto (RIO), Royal Dutch Shell (SHEL), Unilever (ULVR), Vodafone Group (VOD).

The price of these stocks is measured in \textit{ticks}\footnote{A \textit{tick} is the minimum change of the price.}. The time is measured in seconds. We choose to adopt the physical time because we believe that investors perceive this one. We checked that the results are different as we analyze the data with the time in ticks or in seconds. In addition to this a measure of the time in \textit{ticks} would make difficult to compare and aggregate the results for different stocks. In fact, while the number of seconds of trading does not change from stock to stock the number of operation per day can be very different. 

We measure the conditional probability of bounce $p(b|b_{prev})$ given $b_{prev}$ previous bounces. This is the probability that the price bounces on a local maximum or minimum given $b_{prev}$ previous bounces. 
Practically, we record if the price, when is within the stripe of a support or resistance, bounces or crosses it for every day of trading and for every stock. We assume that all the supports or resistances detected in different days of the considered year are statistically equal. As a result we obtain the bounce frequency $f(b|b_{prev})=\frac{n_{b_{prev}}}{N}$ for the total year. Now we can estimate $p(b|b_{prev})$ with the method of the Bayesian inference: we infer $p(b|b_{prev})$ from the number of bounces $n_{b_{prev}}$ and from the total number of trials $N$ assuming that $n$ is a realization of a Bernoulli process because when the price is contained into the stripe of a previous local minimum or maximum it can only bounce on it or cross it.

\begin{figure}[h!]

\begin{tabular}{cc}
\epsfig{file=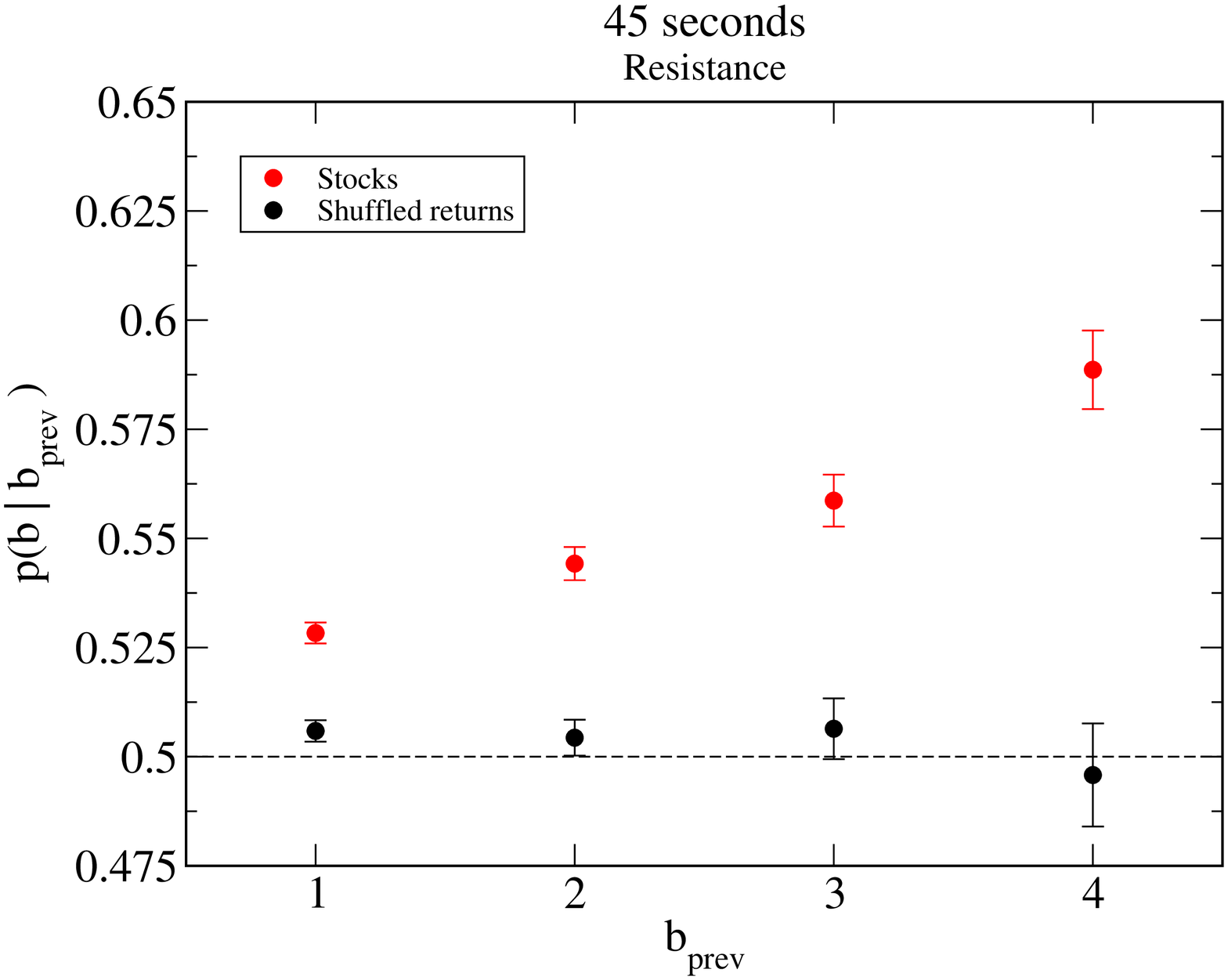,width=0.55\linewidth,clip=} &
\epsfig{file=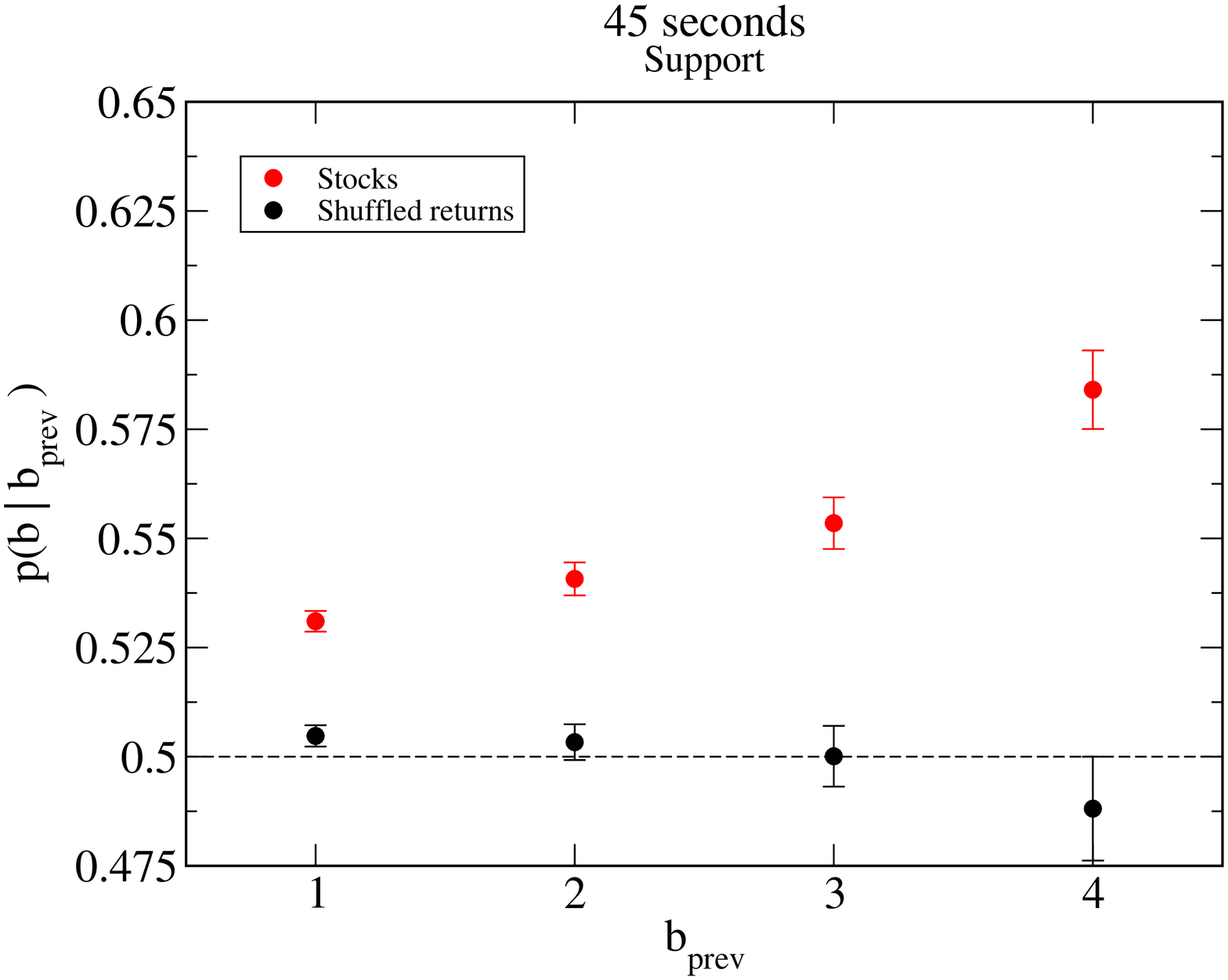,width=0.55\linewidth,clip=} \\
\epsfig{file=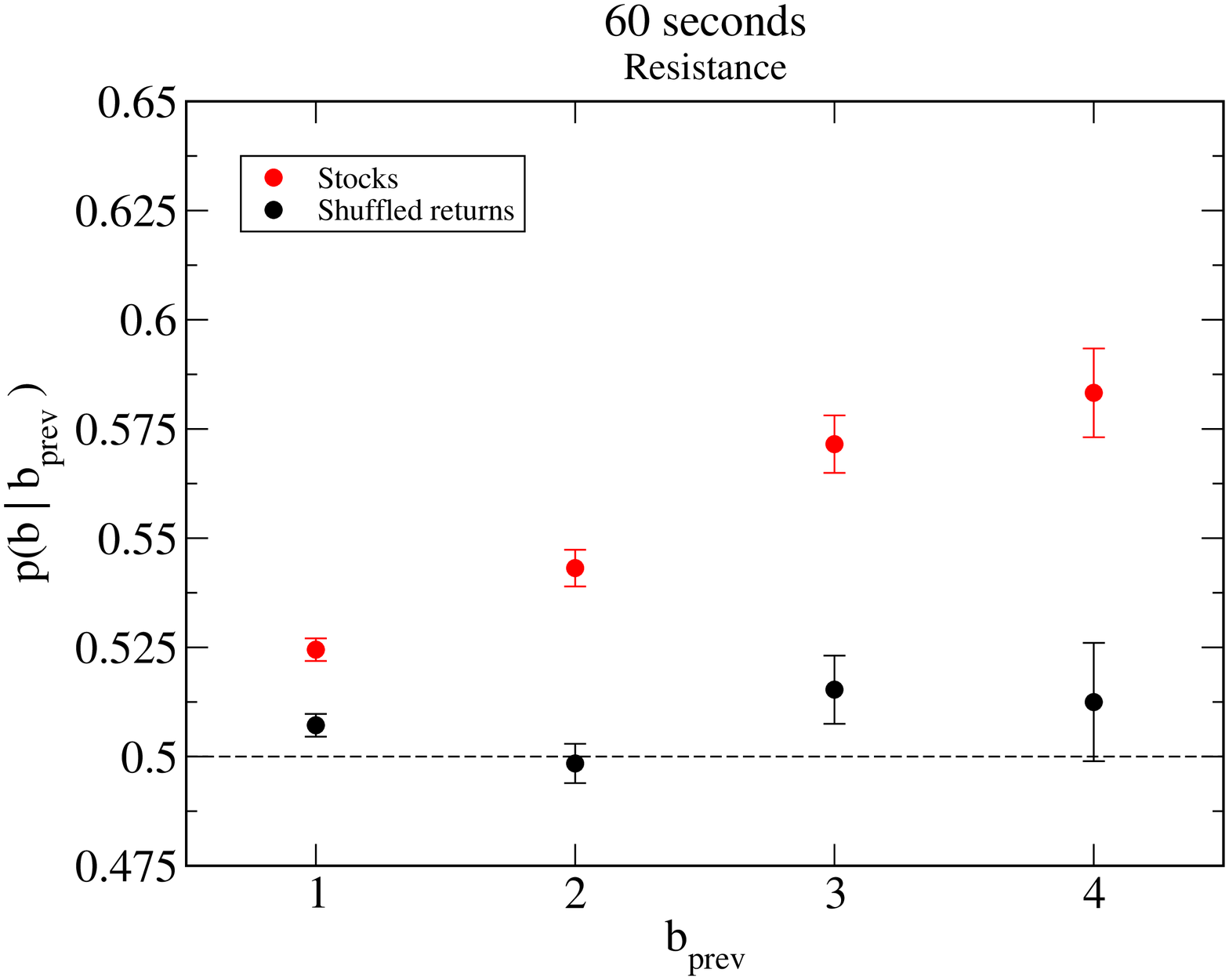,width=0.55\linewidth,clip=} &
\epsfig{file=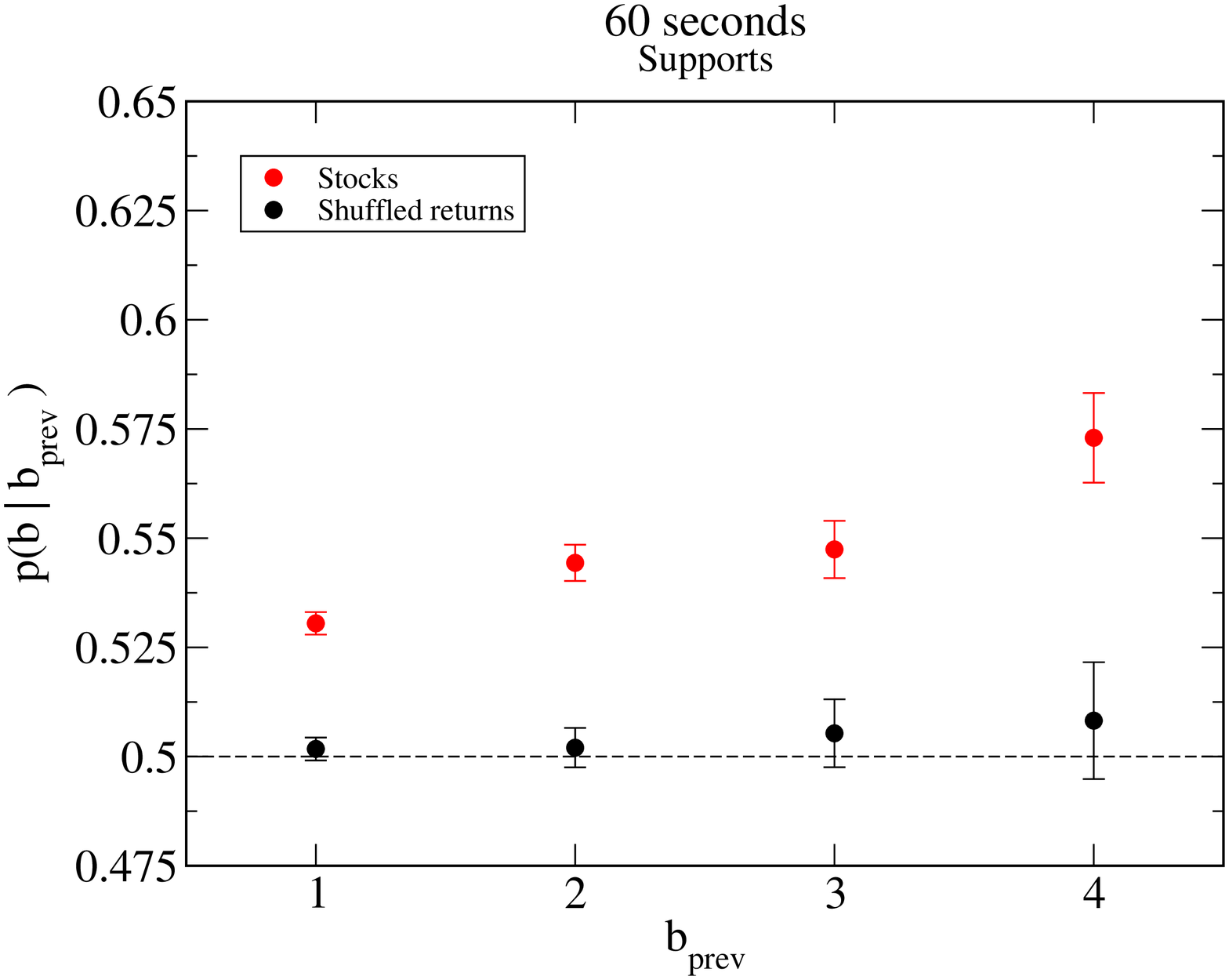,width=0.55\linewidth,clip=} \\
\end{tabular}
\caption{Graphs of the conditional probability of bounce on a resistance/support given the occurrence of $b_{prev}$ previous bounces. Time scale: $\tau$=45, 60 seconds. The data refers to the 9 stocks considered. The data of the stocks are shown as red circles while the data of the time series of the shuffled returns of the price are shown as black circles. The graphs in the left refer to the resistances while the ones on the right refer to the supports. }
\label{fig:timeseconds1}
\end{figure}

\begin{figure}[h!]
\centering
\begin{tabular}{cc}
\epsfig{file=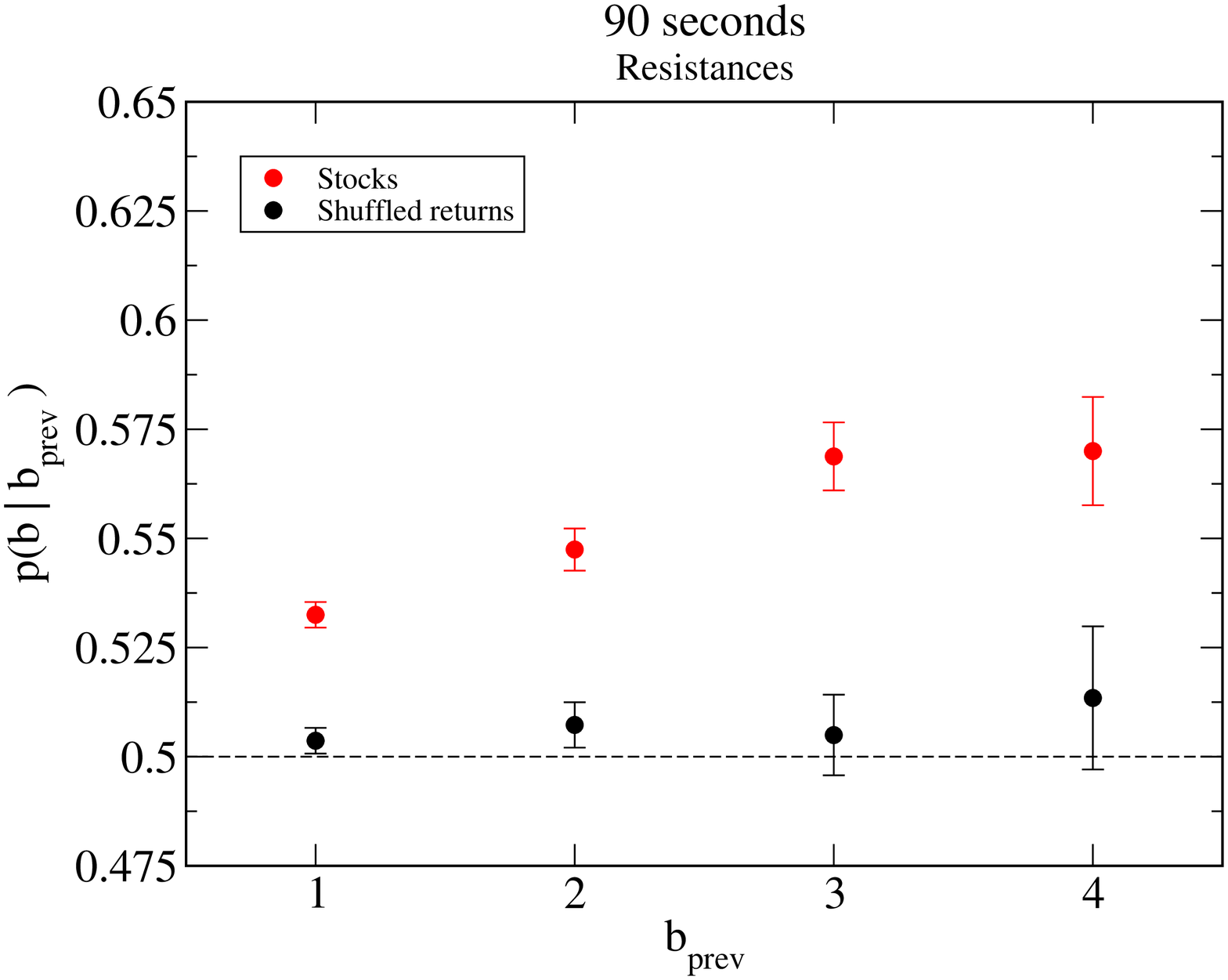,width=0.55\linewidth,clip=} &
\epsfig{file=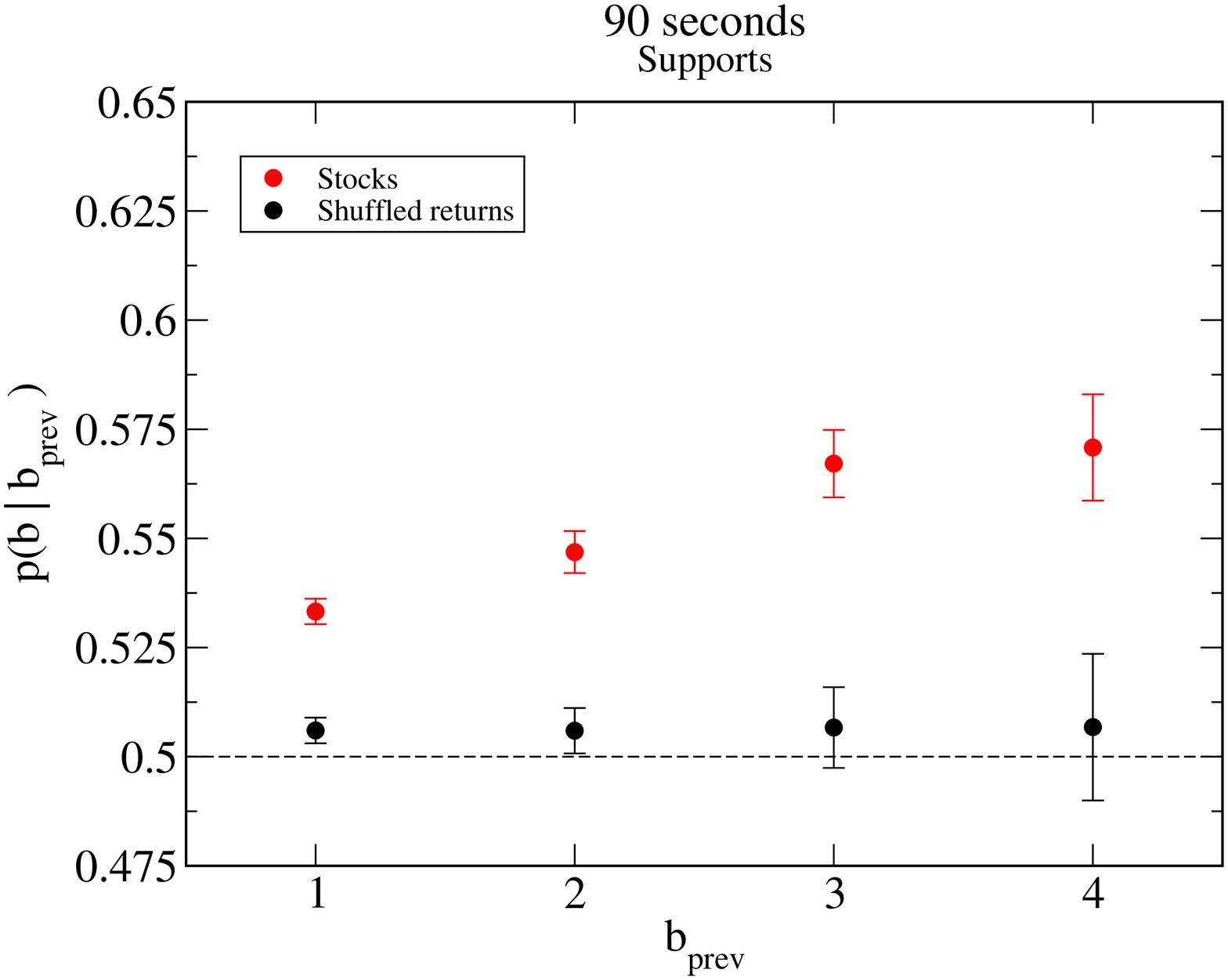,width=0.55\linewidth,clip=} \\
\epsfig{file=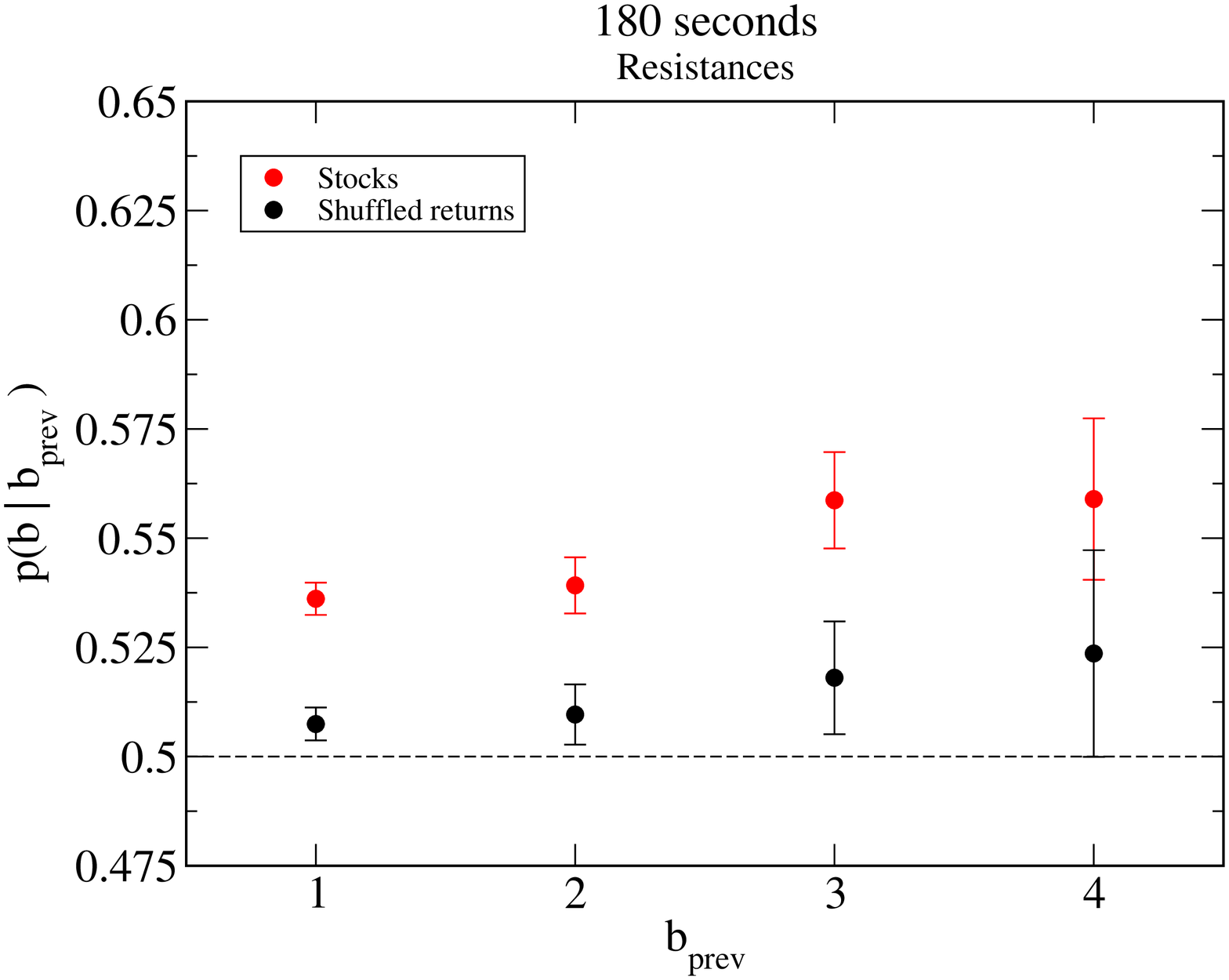,width=0.55\linewidth,clip=} &
\epsfig{file=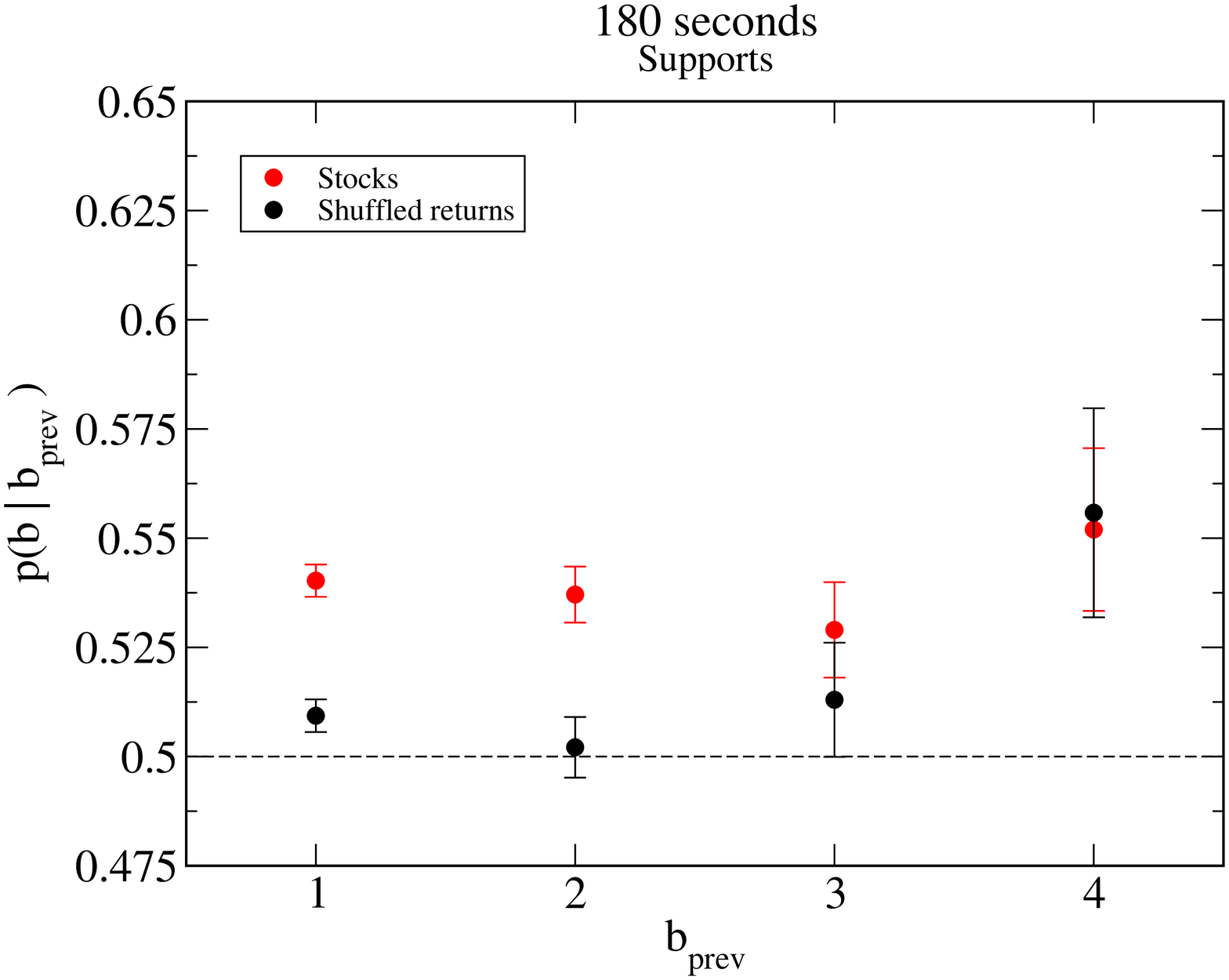,width=0.55\linewidth,clip=}
\end{tabular}
\caption{Graphs of the conditional probability of bounce on a resistance/support given the occurrence of $k$ previous bounces. Time scale: $\tau$=90, 180 seconds. The data refers to the 9 stocks considered. The data of the stocks are shown as red circles while the data of the time series of the shuffled returns of the price are shown as black circles. The graphs in the left refer to the resistances while the ones on the right refer to the supports.}
\label{fig:timeseconds2}
\end{figure}

Using this framework we can evaluate the expected value and the variance of $p(b|b_{prev})$ using the Bayes theorem \cite{feller, D'Agostini}:
\begin{eqnarray}
E[p(b|b_{prev})]&=&\frac{n_{b_{prev}}+1}{N+2}\\
Var[p(b|b_{prev})]&=&\frac{(n_{b_{prev}}+1)(N-n_{b_{prev}}+1)}{(N+3)(N+2)^2} \label{eq:varbayes}
\end{eqnarray}
In fig.\ref{fig:timeseconds1} and fig.\ref{fig:timeseconds2} the conditional probabilities are shown for different time scales. The data of the stocks have been compared to the time series of the shuffled returns of the price. In this way we can compare the stock data with a time series with the same statistical properties but without any memory effect. As shown in the graphs, the probabilities of bounce of the shuffled time series are nearly $0.5$ while the probabilities of bounce of the stock data are well above 0.5. In addition to this, it is noticeable that the probability of bounce rises up as $b_{prev}$ increases. Conversely, the probability of bounce of the shuffled time series is nearly constant. The increase of $p(b|b_{prev})$ of the stocks with $b_{prev}$ can be interpreted as the growth of the investors' trust on the support or the resistance as the number of bounces grows. The more the number of previous bounces on a certain price level the stronger the trust that the support or the resistance cannot be broken soon. As we outlined above, a feedback effect holds therefore an increase of the investors' trust on a support or a resistance entails a decrease of the probability of crossing that level of price.

We have performed a $\chi^2$ test to verify if the hypothesis of growth of $p(b|b_{prev})$ is statistically meaningful. The independence test ( $p(b|b_{prev})=c$ ) is performed both on the stock data and on the data of the shuffled time series and we compute $$\chi^2=\frac{\sum_{b_{prev}=1}^{4}\left[p(b|b_{prev})-c\right]^2}{\sum_{b_{prev}=1}^{4}\sigma_{b_{prev}}^{2}}\,\mbox{.}$$ Then we compute the p-value associated to a $\chi^2$ distribution with 2 degrees of freedom. We choose a significance level $\alpha=0.05$. If $\text{p-value}<\alpha$ the independence hypothesis is rejected while if  $\text{p-value} \geqslant \alpha$ it is accepted. The results are shown in table \ref{table_p-value}. The green cells indicate independence of $p(b|b_{prev})$ on the value of $b_{prev}$ while the red cells indicate dependence of $p(b|b_{prev})$. 
The results show that there is a clear increase of the investors' memory on the supports/resistances as the number of previous bounces increases for the time scales of 45, 60 and 90 seconds. Conversely, this memory do not increase at the time scale of 180 seconds.  

\begin{table}[h!]
\begin{tabular}{|cl||c|c|c|c||}
\hline		
\hline
 						& & 45 sec. & 60 sec. & 90 sec. & 180 sec. \\
\hline
\multirow{2}*{Stocks}		
        
   					 & resistances & \cellcolor{light-red} $< 0.0001$ & \cellcolor{light-red} $< 0.0001$ & \cellcolor{light-red} $< 0.0001$ & \cellcolor{light-green} 0.077 \\
        		     & supports & \cellcolor{light-red} $< 0.0001$ & \cellcolor{light-red} $< 0.0001$ & \cellcolor{light-red} $< 0.0001$ & \cellcolor{light-green} 0.318 \\ 
\hline
\hline
\multirow{2}*{Shuffled returns} & resistances & \cellcolor{light-green} 0.280 & \cellcolor{light-green} 0.051 & \cellcolor{light-green} 0.229
 &  \cellcolor{light-green} 0.583 \\
        						  & supports & \cellcolor{light-green}  0.192 & \cellcolor{light-green} 0.229 & \cellcolor{light-green} 0.818 & \cellcolor{light-green} 0.085 \\	
\hline				
\hline

\end{tabular}
\caption{The table shows the p-values for the stock data and for the time series of the shuffled returns for different time scale and for the supports and resistances. The red cells indicate independence of $p(b|b_{prev})$ to the $b_{prev}$ value. The green cells indicate a non trivial dependence of $p(b|b_{prev})$ to the $b_{prev}$ value.}
\label{table_p-value}
\end{table}

\section{Long memory of the price}

The analysis of the conditional probability $p(b|b_{prev})$ proves the existence of a long memory in the price time series. We used the Hurst exponent $H$ as a measure of a such long term memory or autocorrelation. The Hurst exponent is estimated via the detrended fluctuation method \cite{peng, heart}. It is useful to recall that the Hurst exponent provides about the autocorrelation of the time series:
\begin{itemize}
\item if $H<0.5$ one has negative correlation and antipersistent behavior
\item if $H=0.5$ one has no correlation
\item if $H>0.5$ one has positive correlation and persistent behavior
\end{itemize}

\begin{figure}[p]
\centering
\subfloat[][Graph of $\ln \sigma(n)$ against $n$ of a trading day of RIO. The red line is a linear fit of the data. Its slope gives an estimation of the Hurst exponent. The black line represents the linear fit that one would obtain for an uncorrelated time series.]
{\includegraphics[width=.45\columnwidth]{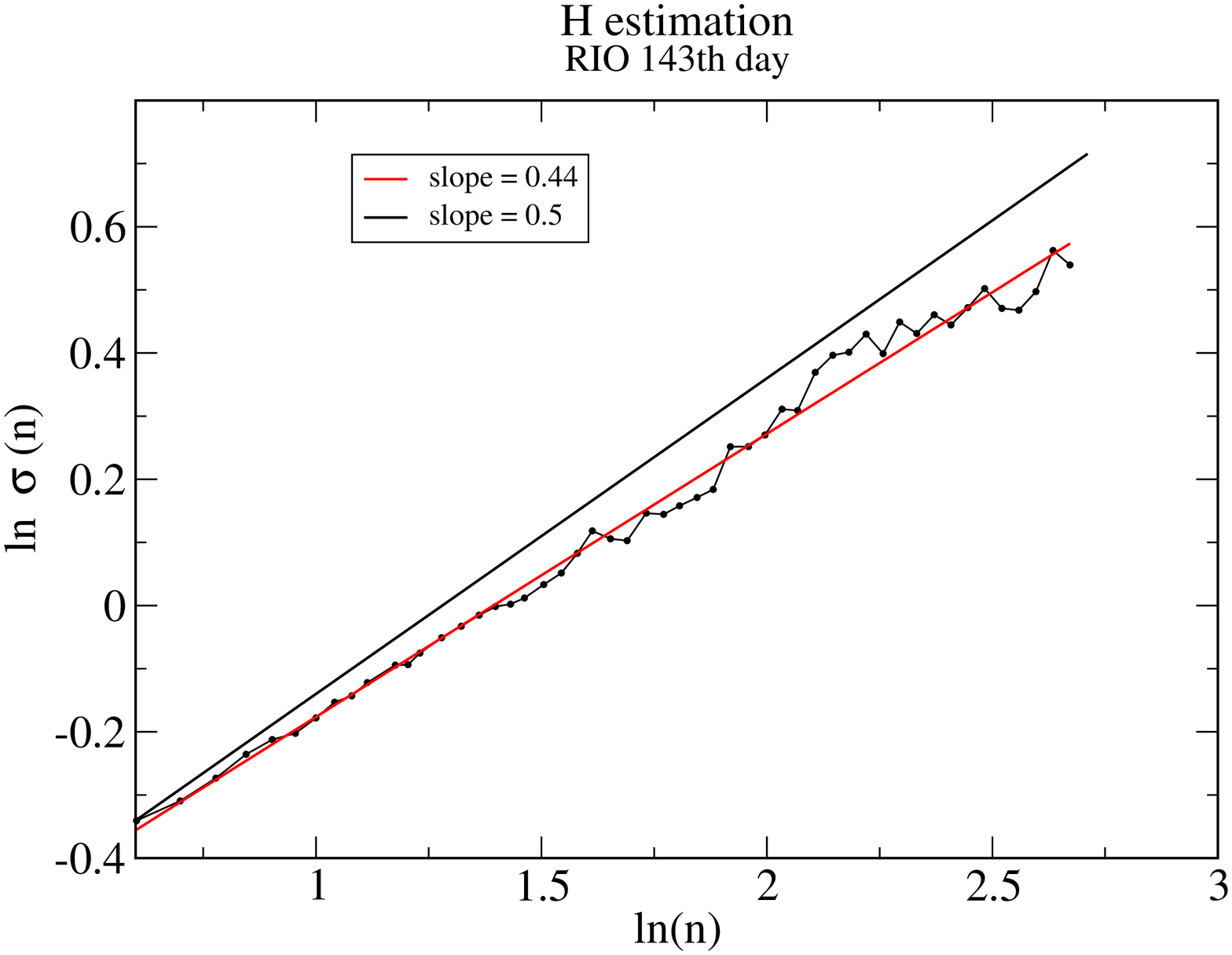}} \quad
\subfloat[][Histogram of $H$ measured in the 251 trading days of 2002 for RIO. The dotted line is the average Hurst exponent over the year 2002 while the red line indicates $H = 0.5$]
{\includegraphics[width=.45\columnwidth]{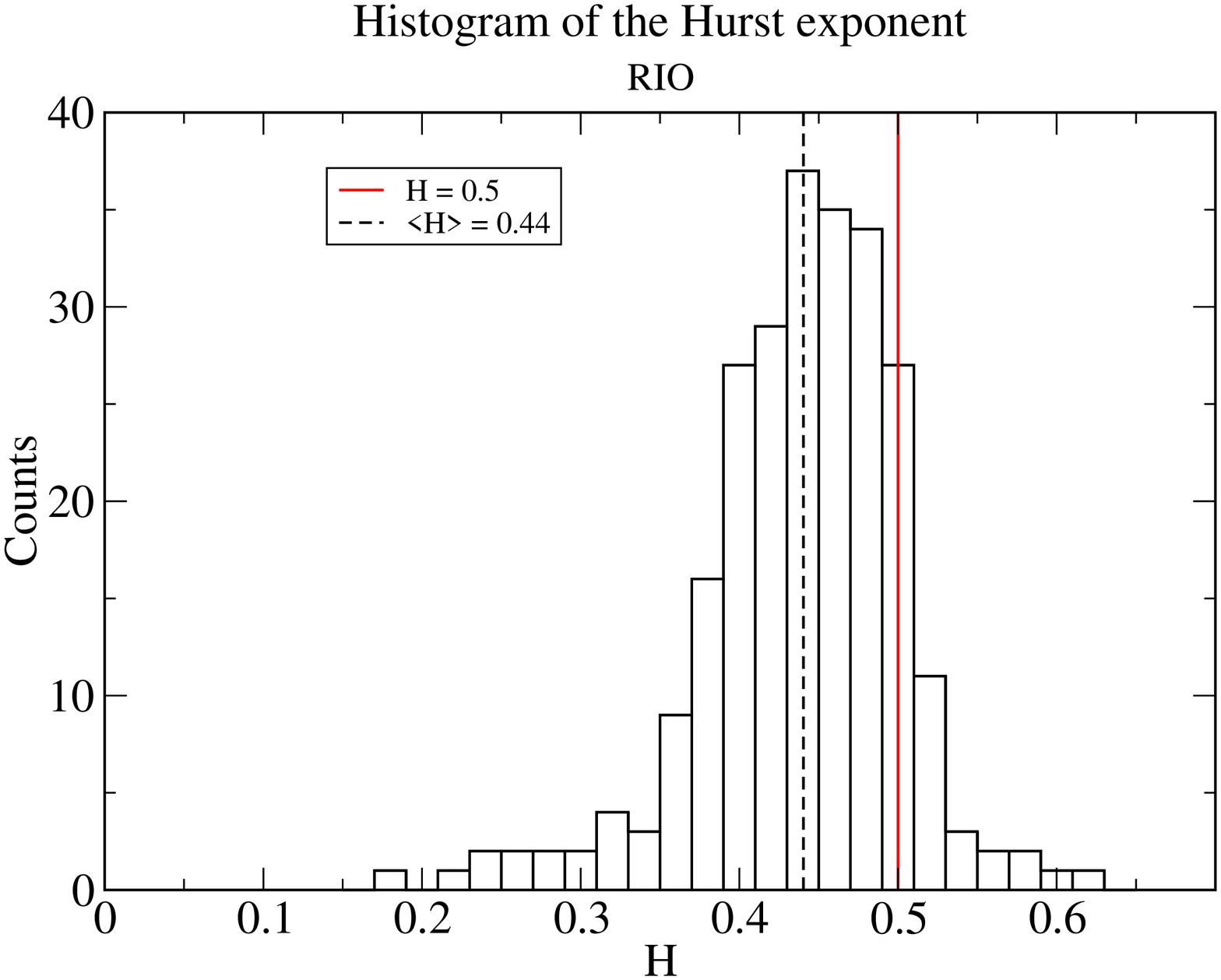}} \\
\caption{}
\label{fig:subdiffusion}
\end{figure}
\subsection{Empirical evidence of the anticorrelation}
If we consider the graph in figure \ref{fig:subdiffusion}(a) it is noticeable that the price increments are anticorrelated in the given day being the slope of the linear fit $H=0.44 < 0.5$. 
The process is not anticorrelated every day: we find $H > 0.5$ in some days and $H < 0.5$ in others. However the average Hurst exponent is $\langle H \rangle < 0.5$ therefore the price increments are anticorrelated on average. The graph  \ref{fig:subdiffusion}(b) shows the histogram of $H$ measured in the 251 trading days of 2002 for RIO. 
The table \ref{tab:diffusion} shows the average values of the Hurst exponent for all the 9 stocks analyzed in this paper.
\begin{table}
\begin{center}
\begin{tabular}{||c|c||} 
\hline Stock &  $H$ \\ 
\hline
\hline AZN & 0.471 \\ 
 BP & 0.440 \\ 
 GSK & 0.464 \\ 
 HBOS & 0.472 \\ 
 RBS & 0.483 \\
 RIO & 0.440 \\
 SHEL & 0.445 \\ 
 ULVR & 0.419 \\
 VOD &  0.419 \\    
\hline 
\end{tabular} 
\end{center}
\caption{Average values of the Hurst exponent over the year 2002 for all the 9 stocks analyzed in this paper.}
\label{tab:diffusion}
\end{table}
The table shows that $\langle H \rangle$ is always less than $0.5$ therefore there is anticorrelation effect of the price increments for the 9 stocks analyzed.

\begin{figure}[p]
\centering
\includegraphics[scale=0.4]{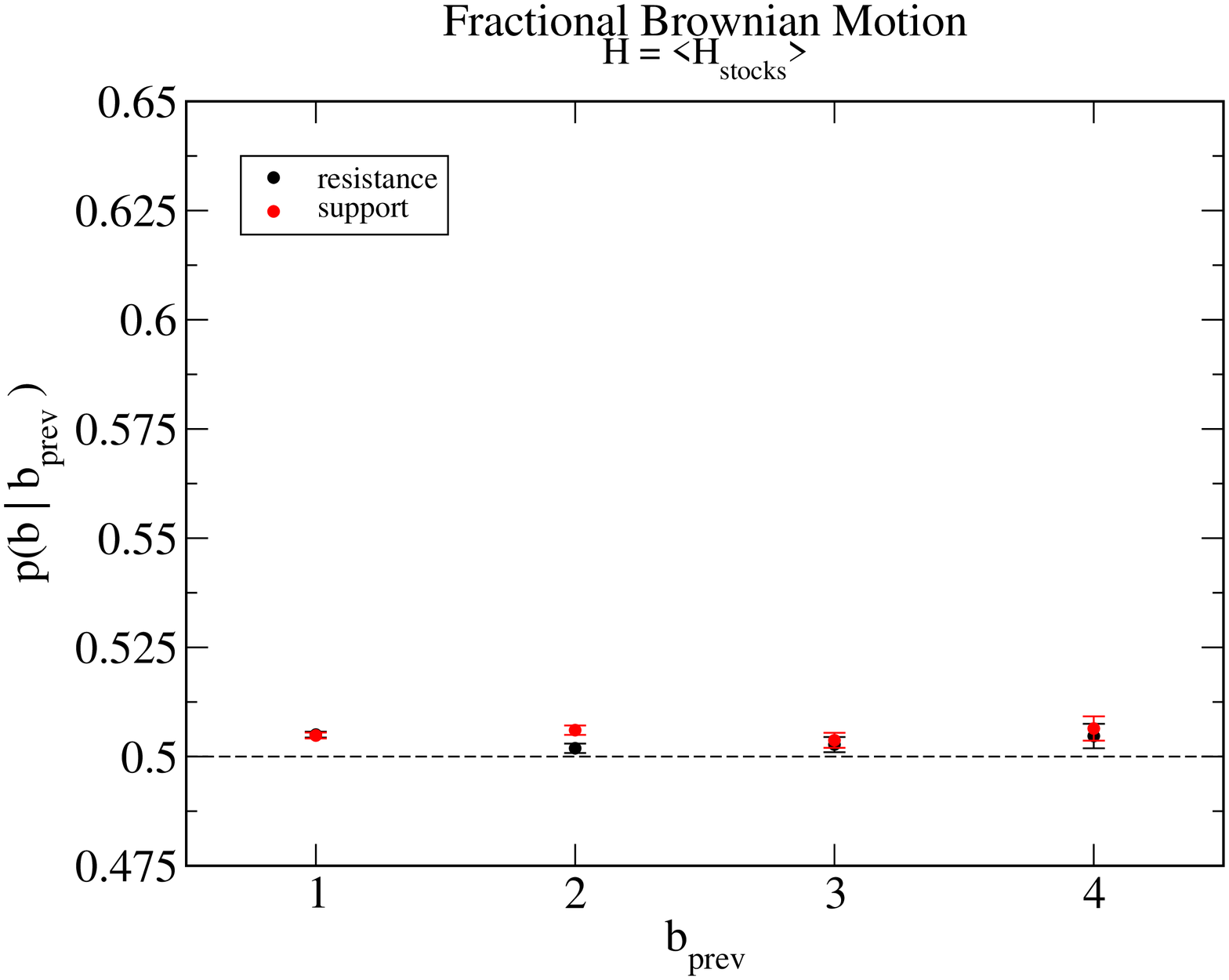} 
\caption{Graph of the conditional probability of bounce on a resistance/support given the occurrence of $b_{prev}$ previous bounces for a fractional random walk. The red circles refers to supports, the black ones to resistances.}
\label{fig:fbmHstocks}
\end{figure}

The anticorrelation of the price increments could lead to an increase of the bounces and therefore it could mimic a memory of the price on a support or resistance. We perform an analysis of the bounces on a antipersistent fractional random walk to verify if the memory effect depends on the antipersistent nature of the price in the time scale of the day. We choose a fractional random walk with the Hurst exponent $H=\langle H_{stock} \rangle$ given by the average over the $H$ exponents of the different stocks shown in table \ref{tab:diffusion}. The result is shown in fig. \ref{fig:fbmHstocks}. The conditional probabilities $p(b|b_{prev})$ are very close to 0.5 although above this value. In addition to this it is clear that  $p(b|b_{prev})$ is constant. These two results prove that the memory effect of the price does not depend on its antipersistent properties, or at least the antipersistence is not the main source of this effect.

\section{Features of the bounces}

\begin{figure}
\centering
\includegraphics[scale=0.5]{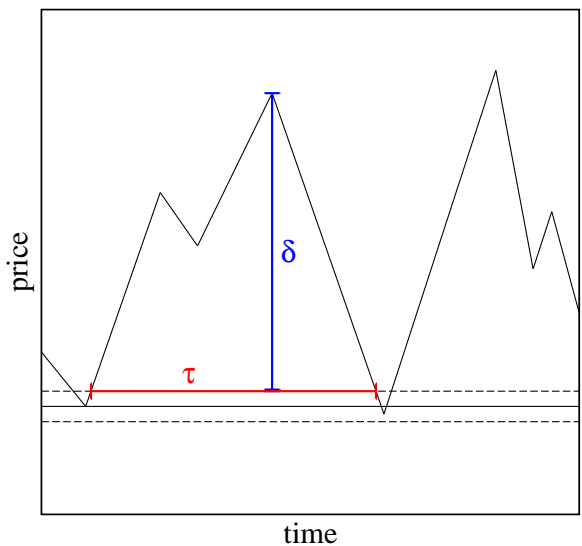} 
\caption{Sketch of the price showing how we defined $\tau$, the time between two bounces and $\delta$, the maximum distance between the price and the support or resistance level between two bounces.}
\label{fig:tau_sketch}
\end{figure}

Now we want to describe two features of the bounces: the time $\tau$ occurring between two consecutive bounces and the maximum distance $\delta$ of the price from the support or the resistance between two consecutive bounces. 

\begin{figure}[h!]
\centering
\begin{tabular}{cc}
\epsfig{file=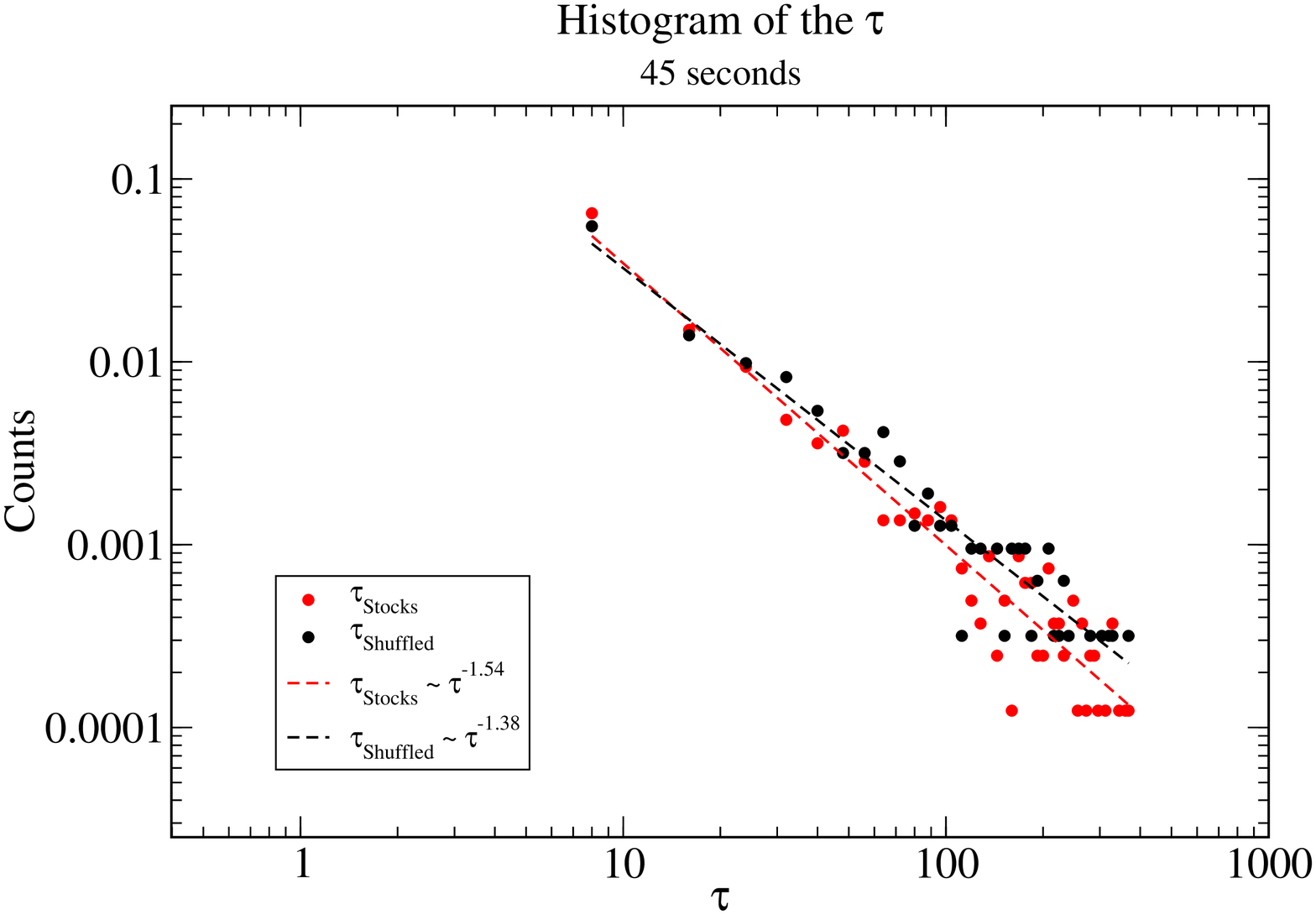,width=0.55\linewidth,clip=} &
\epsfig{file=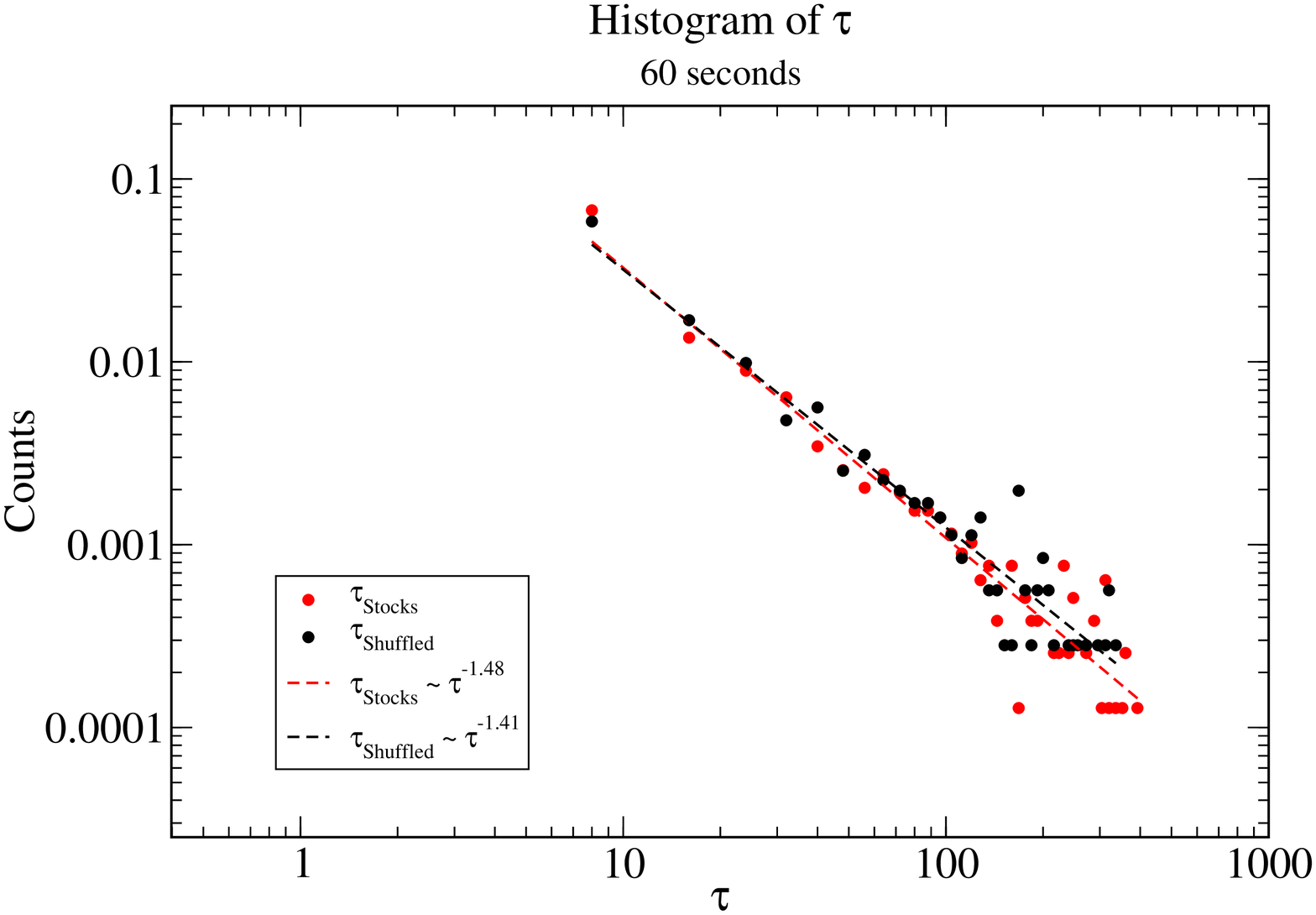,width=0.55\linewidth,clip=} \\
\epsfig{file=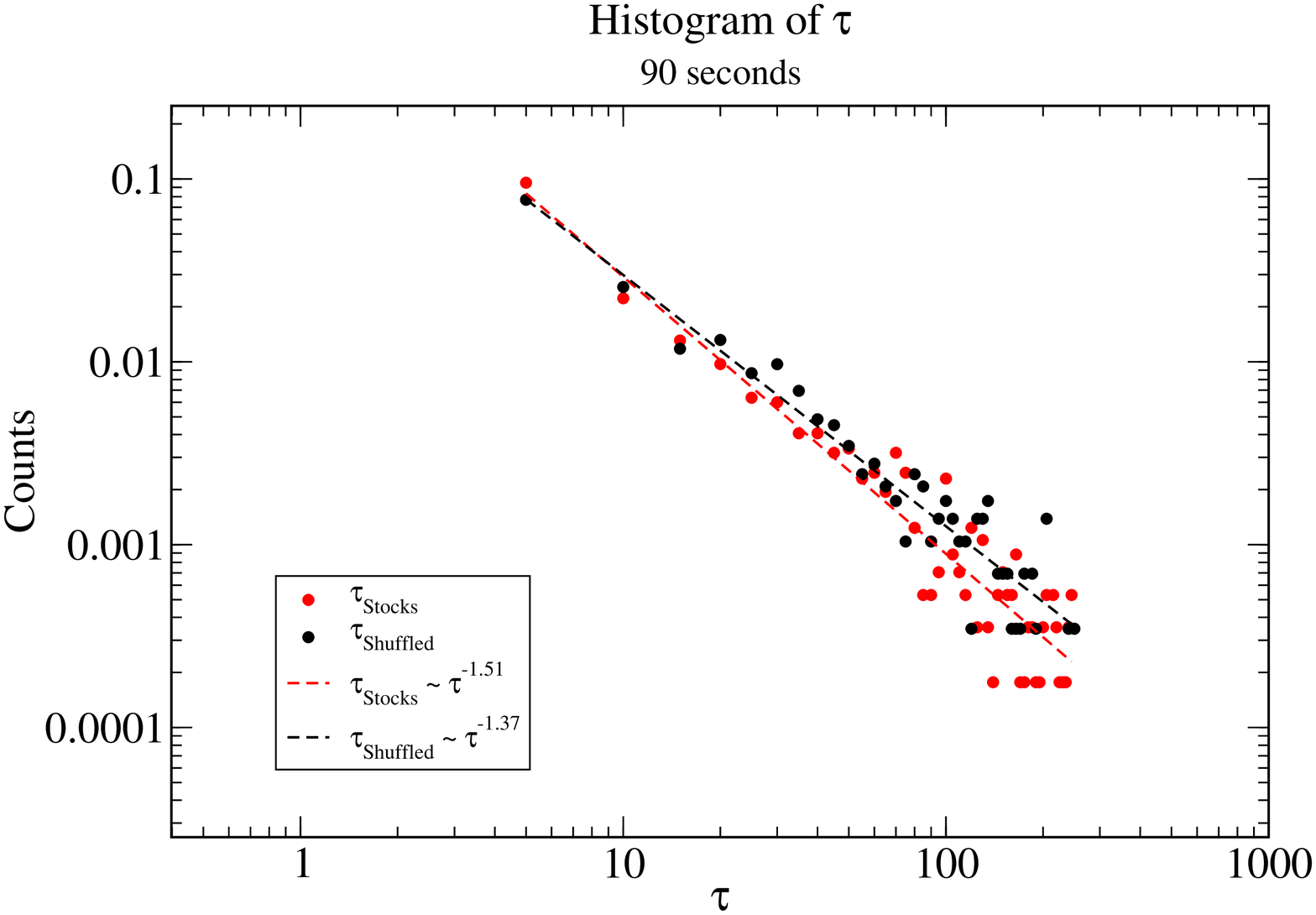,width=0.55\linewidth,clip=} &
\epsfig{file=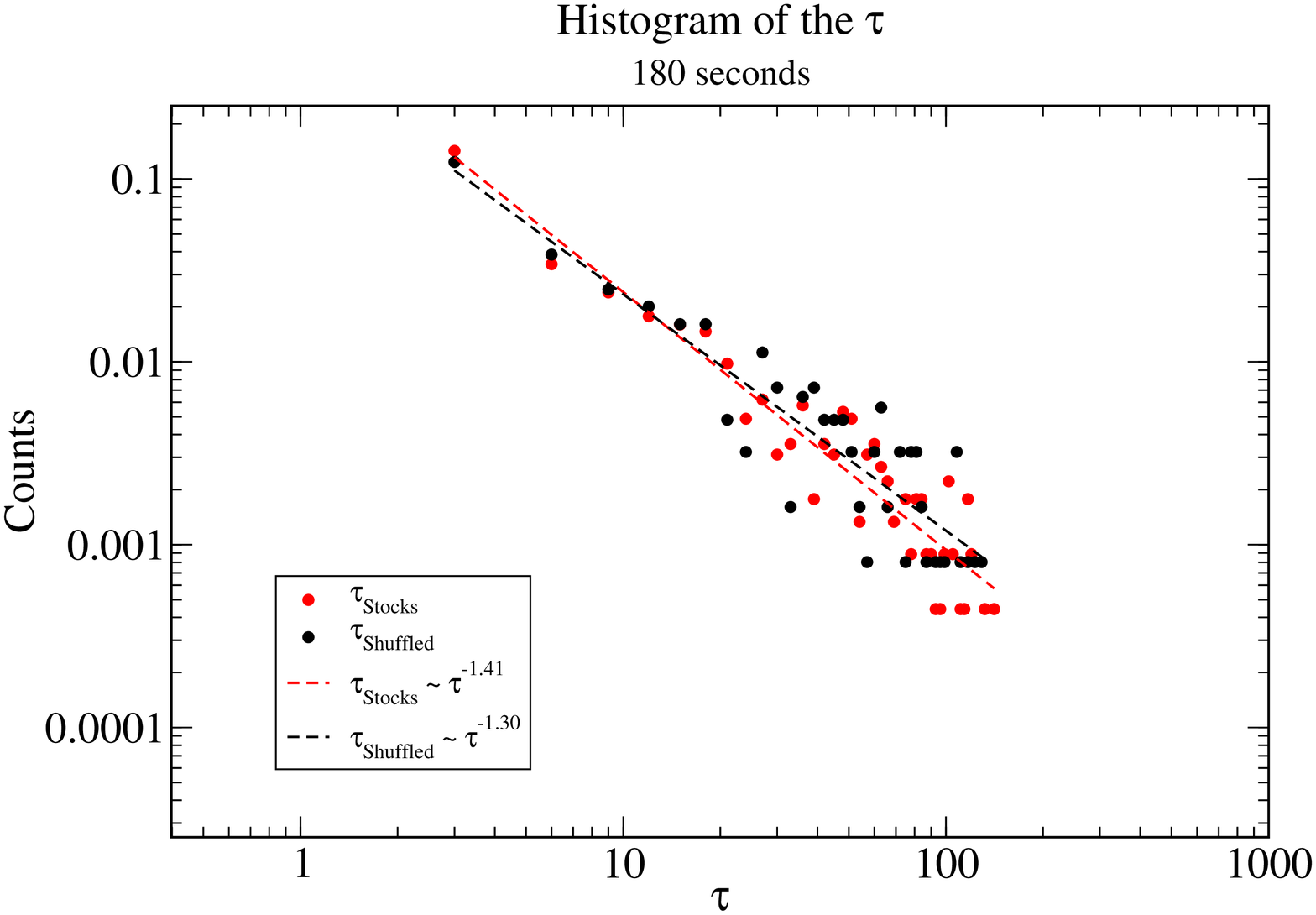,width=0.55\linewidth,clip=}
\end{tabular}
\caption{Graphs of the histograms of $\tau$ for the time scales of 45, 60, 90, 180 seconds. We obtained the histograms from the data of all the 9 stocks analyzed in this paper. We do not make any difference between supports and resistances in this analysis. The red circles are related to the stocks while the black ones are related to the shuffled time series. The red dotted line is a power law fit of the stocks data, the black dotted line is a power law fit of the shuffled time series data. The time $\tau$ is measured in terms of the considered time scale.}
\label{fig:tau}
\end{figure}

The time of recurrence $\tau$ is defined as the time between an exit of the price from the stripe centered on the support or resistance and the return of the price in the same stripe, as shown in fig.\ref{fig:tau_sketch}. We study the distribution  of $\tau$ for different time scales (45, 60, 90 and 180 seconds) and for the normal and the shuffled time series (fig.\ref{fig:tau}). We find no significant difference between the two histograms. We point out that, being $\tau$ measured in terms of the considered time scale, we can compare the four histograms. We find that a power law fit describes well the histograms of $\tau$.

\begin{figure}
\centering
\begin{tabular}{cc}
\epsfig{file=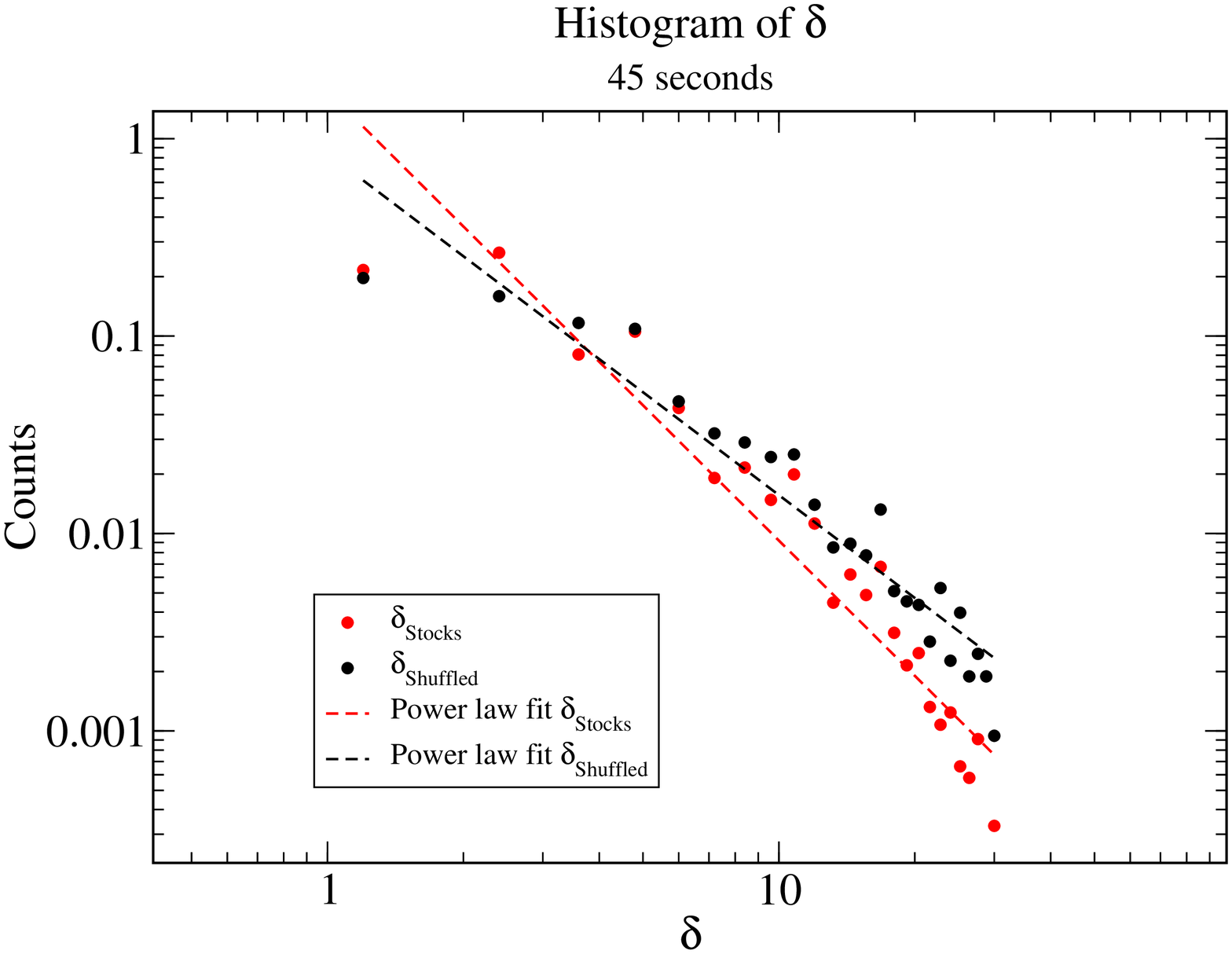,width=0.55\linewidth,clip=} &
\epsfig{file=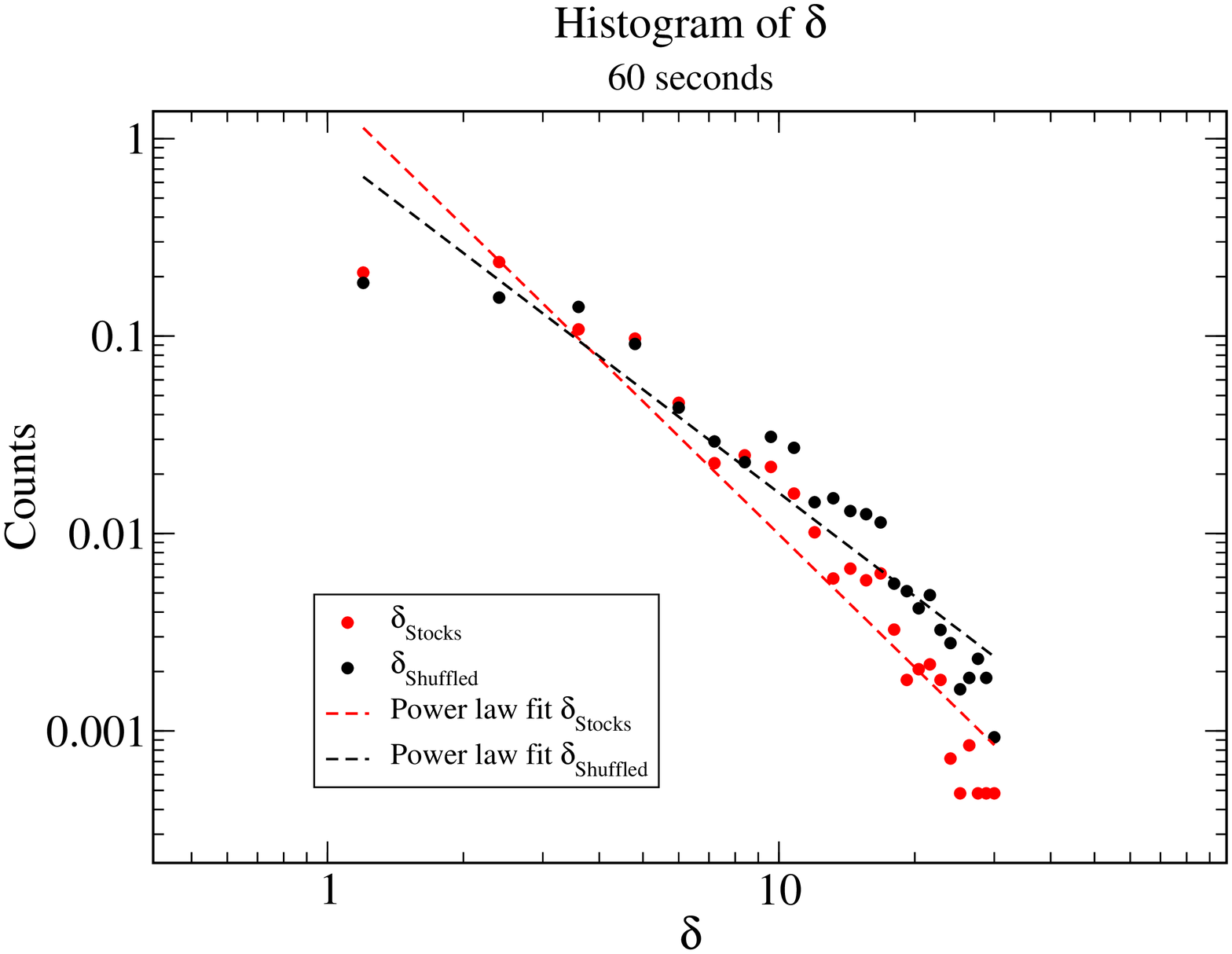,width=0.55\linewidth,clip=} \\
\epsfig{file=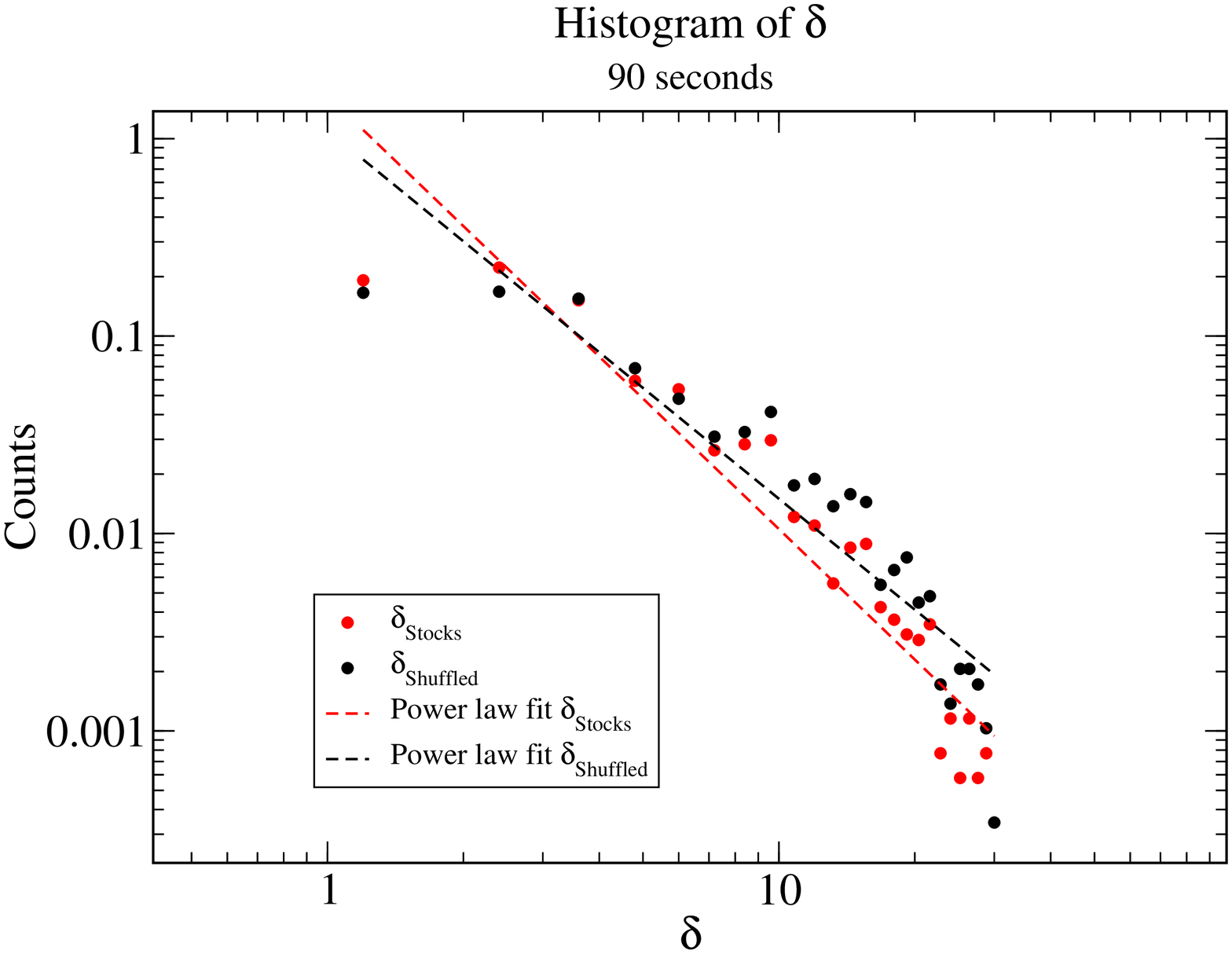,width=0.55\linewidth,clip=} &
\epsfig{file=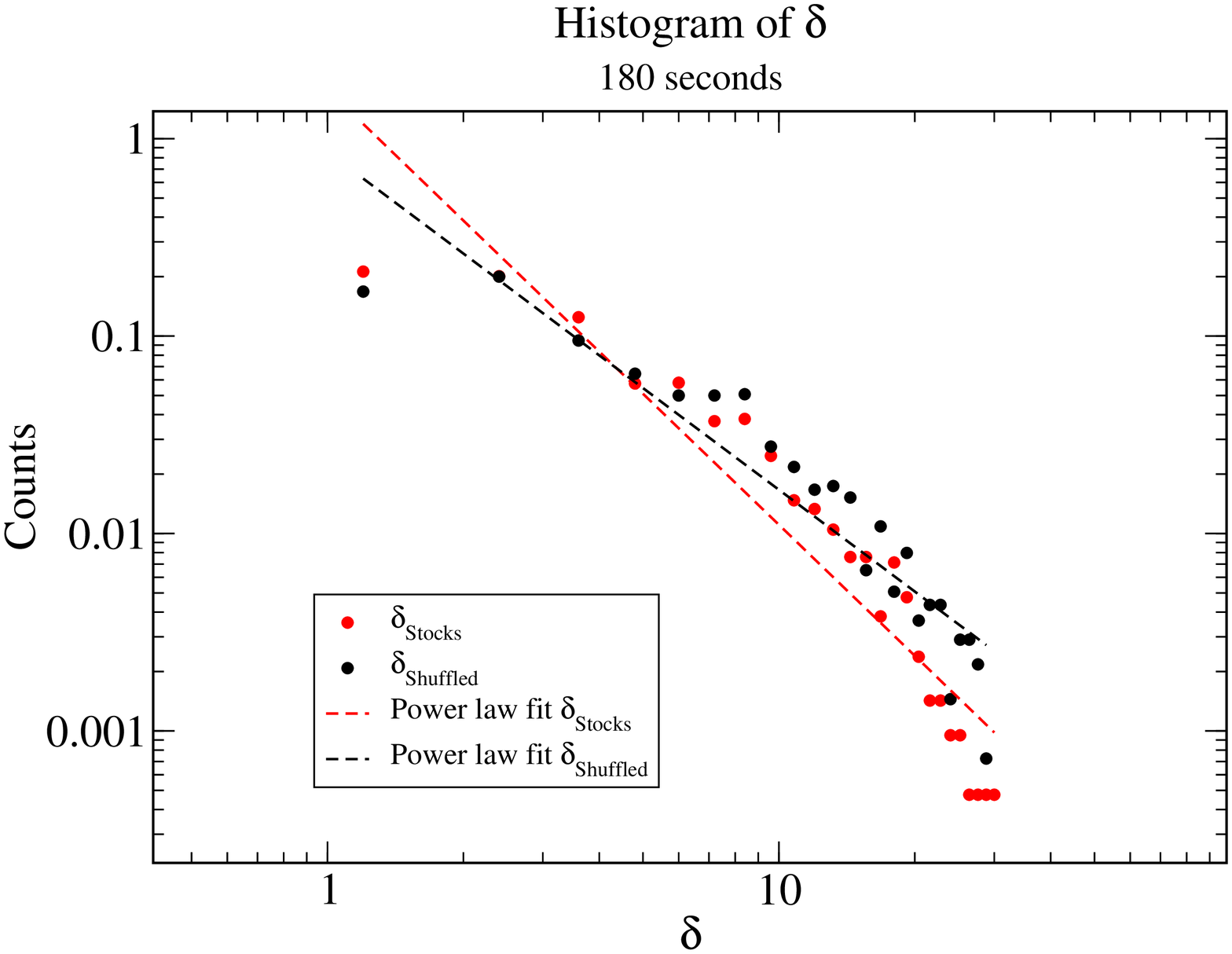,width=0.55\linewidth,clip=}
\end{tabular}
\caption{Graphs of $\delta$ for the time scales of 45, 60, 90, 180 seconds. We obtained the histograms from the data of all the 9 stocks analyzed in this paper. We do not make any difference between supports and resistances in this analysis. The red circles are related to the stocks while the black ones are related to the shuffled time series. The red dotted line is a power law fit of the stocks data, the black dotted line is a power law fit of the shuffled time series data. The price $\delta$ is measured in ticks.}
\label{fig:delta}
\end{figure}

We call $\delta$ the maximum distance reached by the price before the next bounce. We show in fig.\ref{fig:tau_sketch}  how the maximum distance $\delta$ is defined. We study the distribution of $\delta$ for different time scales (45, 60, 90 and 180 seconds) and for the normal and the shuffled time series is shown in fig.\ref{fig:delta}. In this case a power law fit does not describe accurately the histogram of $\delta$. The histograms have a similar shape, but the tail of the distributions of the stock data is thinner for both quantities $\tau$ and $\delta$. According to this two evidences the memory effect of the price due to supports and resistances leads to smaller and more frequent bounces.

\section{Conclusion}
In this paper we perform a statistical analysis of the price dynamics of several stocks traded at the London Stock Exchange 
in order to verify if there exists an empirical and detectable evidence for technical trading. 
In fact it is known that there exist investors which use technical analysis as trading strategy (also known as chartists). 
The actions of this type of agents may introduce a feedback effect, the so called \textit{self-fulfilling prophecy}, 
which can lead to detectable signals and in some cases to 
arbitrage opportunities, in details, these feedbacks can introduce some memory effects on the price evolution.
The main goal of this paper is to determine if such memory in price dynamics exists or not and consequently if 
it it possible to quantify the feedback on the price dynamics deriving from some types of technical trading strategies.
In particular we focus our attention on a particular figure of the technical analysis called \textit{supports and resistances}.
In order to estimate the impact on the price dynamics of such a strategy 
we measure the conditional probability of the bounces $p(b|b_{prev})$ given $b_{prev}$ previous bounces on a set of suitably selected price values
 to quantify the memory introduced by the supports and resistances. \\
We find that the probability of bouncing on support and resistance
values is higher than $1/2$ or, anyway, is higher than an equivalent random walk or of
the shuffled series. In particular we find that as the number of bounces on
these values increases, the probability of bouncing on them becomes higher. This
means that the probability of bouncing on a support or a resistance is an increasing
function of the number of previous bounces, differently from a random walk or from
the shuffled time series in which this probability is independent on the number of
previous bounces. This features is a very interesting quantitative evidence for a self-reinforcement of
agents' beliefs, in this case, of the strength of the resistance/support.

As a side result we also develop a criterion to select the price values 
that can be potential supports or resistances on which the probability of the bounces $p(b|b_{prev})$ is measured.

We point out that this finding is, in principle, an arbitrage opportunity
because, once the support or the resistance is detected, the next time the price will
be in the nearby of the value a re-bounce will be more likely than the crossing of the
resistance/support. However, when transaction costs and frictions (i.e.
the delay between order submissions and executions) are taken into account, these
minor arbitrage opportunities are usually no more profitable. 

\section*{Acknowledgements}
This research was partly supported by EU Grant FET Open Project 255987 FOC.


\end{document}